# Anomalous density fluctuations in a strange metal

M. Mitrano[a,1], A. A. Husain[a], S. Vig[a], A. Kogar[a,b], M. S. Rak[a], S. I. Rubeck[a], J. Schmalian[c], B. Uchoa[d], J. Schneeloch[e], R. Zhong[e], G. D. Gu[e], P. Abbamonte[a,1]

[a] Department of Physics and Materials Research Laboratory, University of Illinois, Urbana, IL 61801, USA; [b] Department of Physics, Massachusetts Institute of Technology, Cambridge, MA 02139, USA; [c] Institute for the Theory of Condensed Matter, Karlsruhe Institute of Technology (KIT), 76131 Karlsruhe, Germany; [d] Department of Physics and Astronomy, University of Oklahoma, Norman, OK 73069, USA; and [e] Condensed Matter Physics and Materials Science Department, Brookhaven National Laboratory, Upton, NY 11973, USA.

[1] To whom correspondence should be addressed. Email: mmitrano@illinois.edu or abbamonte@mrl.illinois.edu.

**Abstract**

**A central mystery in high temperature superconductivity is the origin of the so-called "strange metal," i.e., the anomalous conductor from which superconductivity emerges at low temperature. Measuring the dynamic charge response of the copper-oxides, $\chi''(q,\omega)$, would directly reveal the collective properties of the strange metal, but it has never been possible to measure this quantity with meV resolution. Here, we present the first measurement of $\chi''(q,\omega)$ for a cuprate, optimally doped Bi$_2$Sr$_2$CaCu$_2$O$_{8+x}$ ($T_c$ = 91 K), using momentum-resolved inelastic electron scattering. In the medium energy range 0.1-2 eV relevant to the strange metal, the spectra are dominated by a featureless, temperature- and momentum-independent continuum persisting to the eV energy scale. This continuum displays a simple power law form, exhibiting $q^2$ behavior at low energy and $q^2/\omega^2$ behavior at high energy. Measurements of an overdoped crystal ($T_c$ = 50 K) showed the emergence of a gap-like feature at low temperature, indicating deviation from power law form outside the strange metal regime. Our study suggests the strange metal exhibits a new type of charge dynamics in which excitations are local to such a degree that space and time axes are decoupled.**

**Significance**

The strange metal is a poorly understood state of matter found in a variety of quantum materials, notably both Cu- and Fe-based high temperature superconductors. Strange metals exhibit a non-saturating, T-linear electrical resistivity, seemingly indicating the absence of electron quasiparticles. Using inelastic electron scattering, we report the first momentum-resolved measurement of the dynamic charge susceptibility of a strange metal, optimally doped Bi$_{2.1}$Sr$_{1.9}$CaCu$_2$O$_{8+x}$. We find that it does not exhibit propagating collective modes, such as the plasmon excitation of normal metals, but instead exhibits a featureless continuum lacking either temperature- or momentum-dependence. Our study suggests the defining characteristic of the strange metal is a singular type of charge dynamics of a new kind for which there is no generally accepted theory.

The nonsuperconducting normal state of the high temperature superconductors, usually referred to as the "strange metal," has many properties that cannot be explained by the conventional Landau-Fermi liquid theory of metals (1,2). These include a resistivity that is linear in temperature and exceeds the Mott-Ioffe-Regel limit (2,3,4,5), an in-plane conductivity exhibiting an anomalous power law dependence on frequency (6,7), a magnetoresistance that violates Kohler's rule (8), a quasiparticle damping, $\Sigma''(\omega)$, that is linear in $\omega$ (9,10), and a nuclear relaxation rate that violates the Korringa law (11). These properties, many of which are also observed in Fe- and Ru- based strange metals (2,12,13,14), imply that metallic quasiparticles are either absent or exist only marginally (15).

The main spectroscopic signature of the strange metal is a featureless continuum observed to the highest measurable energy in Raman scattering experiments (16,17). Its origin is still unknown, exemplifying the need for a new experimental probe of the collective excitations of strange metals, particularly one that could determine how this continuum evolves at finite momentum, *q*. Generically, the most direct measure of the collective excitations of any material is its dynamic charge susceptibility, $\chi(q,\omega)$, which reveals bosonic modes such as the plasmon excitations in ordinary metals (18). Unfortunately, it has never been possible to measure this quantity for the cuprates, at least for $q \neq 0$ at the meV scale relevant to these materials (19).

Here we report the first meV-resolved, $q \neq 0$ measurement of the low-energy dynamic charge response of a strange metal, the optimally-doped cuprate $Bi_{2.1}Sr_{1.9}CaCu_2O_{8+x}$ (BSCCO), using high-resolution, momentum-resolved electron energy-loss spectroscopy (M-EELS) (19,20,21). We focus here on the medium energy region, 0.1-2 eV, most relevant to the strange metal physics. Note that this energy scale is large compared to the temperature or the superconducting gap, comparable to the superexchange, *J*, and small compared to the bandwidth or the Hubbard *U*, so reflects the properties of the metallic phase out of which superconductivity forms. M-EELS measurements were performed on cleaved crystals of optimally doped ($T_c$ = 91 K) and overdoped ($T_c$ = 50 K) BSCCO (Fig. 1B) using 50 eV electrons in reflection geometry (Fig. 1A) with the energy resolution set to 4 meV. In this article the momentum, $q = |(q_x, q_y)|$, will be expressed in tetragonal reciprocal lattice units (r.l.u.) with lattice parameter $a = 3.81$ Å.

M-EELS measures the surface density-density correlation function, $S(q,\omega)$, which is related to the imaginary part of the dynamic charge response, $\chi''(q,\omega)$, by the fluctuation-dissipation theorem (19,21). $\chi''(q,\omega)$ was determined from the M-EELS data by antisymmetrizing the spectra, which eliminates the Bose factor, and scaling the overall magnitude according to the *f*-sum rule (19). The latter (Eq. 6) normalizes out minor intensity drifts in the experiment and determines $\chi''(q,\omega)$ in absolute units (see Methods).

Fig. 2A shows $\chi''(q,\omega)$ for optimally-doped BSCCO at $T$ = 295 K for selected momenta along the $(1,\bar{1})$ direction, perpendicular to the structural supermodulation (22). The large signal at energies below 0.1 eV is from phonon excitations reported previously (19). At low-momenta ($q < 0.15$ r.l.u.), the spectra exhibit a plasmon mode at $\omega_p \sim 1$ eV, which was previously reported in many studies (SI Appendix, Fig. S1). Its broad linewidth indicates that this plasmon is overdamped.

As the momentum is increased to beyond $q > 0.15$ r.l.u., the plasmon fades into a featureless, energy-independent continuum resembling that of early Raman studies (16,17). This continuum is extremely strong, comprising > 99% of the total spectral weight in the *f*-sum rule, and is constant up to an energy scale of 1 eV, suggesting it is electronic in origin. The continuum was found to be essentially isotropic in the $(a,b)$ plane (SI Appendix, Fig. S2) and temperature-independent between room temperature and $T$ = 20 K (Fig. 3A). At energies above 1 eV the susceptibility decays like a power law, $\chi'' \sim 1/\omega^2$.

The momentum dependence of $\chi''(q,\omega)$ is highly anomalous (Fig. 2A). While its magnitude grows like $q^2$, which is required to be consistent with the *f*-sum rule (18,19), the shape of the spectrum is momentum-independent from $q = 0.15$ r.l.u. up to the highest momentum studied, $q = 0.5$ r.l.u. This behavior is highly unlike that of a Fermi liquid whose propagating quasiparticles lead to a strongly momentum-dependent susceptibility, as illustrated in Fig. 1C-D.

The broad plasmon linewidth at small momentum is evidence that the continuum is present even for $q < 0.15$ r.l.u., which would lead to decay of the plasmon via Landau damping (18). To evaluate this possibility, we determined the polarizability of the system, $\Pi(q,\omega)$, which is related to the susceptibility by (18)

$$\chi(q,\omega) = \frac{\Pi(q,\omega)}{\varepsilon_\infty - V(q)\Pi(q,\omega)} \; , \qquad (1)$$

where $V(q)$ is the Coulomb interaction and $\varepsilon_\infty$ is the background dielectric constant, equal to 4.5 in this case (23). The denominator of Eq. 1, $\varepsilon(q,\omega) = \varepsilon_\infty - V(q)\Pi(q,\omega)$, may be thought of as the dielectric function of the system. The difference between the polarizability and the susceptibility is that the former excludes the long-ranged part of the Coulomb interaction, revealing the particle-hole excitation spectrum without interference from plasmon effects.

Determining $\Pi(q,\omega)$ from Eq. 1 is complicated by the fact that the functional form of the Coulomb interaction, $V(q)$, is not precisely known. In a homogeneous, three-dimensional system, $V(q) = 4\pi e^2/q^2$, however M-EELS is a surface probe, and other functional forms are possible near a surface, in layered materials like BSCCO, or in the presence of strong screening (24,25).

For this reason, we modeled the particle-hole continuum using the empirical expression (26),

$$\Pi''(q,\omega) = -\Pi_0(q)\tanh\left[\frac{\omega_c^2(q)}{\omega^2}\right] \; . \qquad (2)$$

This function mimics the experimental data at $q > 0.15$ r.l.u., where $\Pi$ and $\chi$ are expected to be equal, interpolating between a constant at low energy and $1/\omega^2$ behavior at high energy. The quantity $\omega_c(q)$ defines the crossover energy between the two regimes and $\Pi_0(q)$ sets the overall magnitude (26). We fit the data for $\omega > 0.1$ eV, i.e., above the phonon features, by Kramers-Kronig transforming Eq. 2 and using $\omega_c(q)$, $\Pi_0(q)$, and $V(q)$ as adjustable parameters. An excellent fit is obtained at all momenta, even those for $q < 0.15$ r.l.u. in which the plasmon peak is present (Fig. 2A-D). The resulting fit values for $V(q)$ (Fig. 2D) have the form of a 2D Coulomb interaction, $V(q) \propto \exp(-qz)/q$, where $q$ is the in-plane momentum and $z \sim 10$ Å, which is consistent with M-EELS being a surface probe. That Eq. 2 fits the data at all momenta suggests that the continuum is present with the functional form of Eq. 2 everywhere in momentum space—not only for $q > 0.15$ r.l.u. but also in the plasmon regime at low momentum.

Having established a plausible form for $V(q)$ (Fig. 2D), we compute an empirical polarizability by multiplying the experimentally measured $\chi''(q,\omega)$ by the fitted function $|\varepsilon(q,\omega)|^2 / \varepsilon_\infty$. Note that the polarizability obtained in this manner is identical to the susceptibility for all $q > 0.15$ r.l.u. Two distinct regimes are observed. The first is illustrated in Fig. 2E which shows the scaled value $\Pi''(q,\omega)/\Pi_0(q)$ against the scaled energy $\omega/\omega_c(q)$ for each experimentally measured momentum value. All the spectra collapse to a single curve, indicating that, at energies below $\omega_c(q)$, the polarizability $\Pi''(q,\omega) \propto q^2$. The second is illustrated in Fig. 2F which shows the unscaled value $\Pi''(q,\omega)$ against the scaled energy $\omega/v_F q$, where $v_F = 2.8$ eV/r.l.u. is the nodal Fermi velocity (27). All the curves collapse again, this time demonstrating that, at energies higher than $\omega_c(q)$, the polarizability $\Pi''(q,\omega) \propto q^2/\omega^2$.

Stated more succinctly, the polarizability $\Pi''(q,\omega)$ has a simple power-law form, exhibiting an energy-independent, $q^2$ form at low energy and $q^2/\omega^2$ form at high energy. The transition between the two regions is defined by the crossover energy, $\omega_c(q)$. Note that a classic Drude response would decay like $1/\omega^3$ at high energy (6), which is required for a convergent sum rule integral, so the observed response is unconventional to the highest energy measured. Moreover, the absence of any dispersing features in the data leads to the surprising conclusion that the collective excitations are completely local, i.e., density fluctuations in space are decoupled from those in time.

The observed power laws might be interpreted as evidence for a quantum critical point (QCP) near optimal doping claimed by many authors (1). To evaluate this possibility, we repeated our experiment on overdoped (OD) BSCCO with $T_c = 50$ K, which is widely believed to exhibit a crossover to a more Fermi liquid-like phase at low temperature (1,5,15). One would expect to observe deviation from simple power law behavior at low temperature. Fig. 3 shows the temperature dependence of the M-EELS spectra from OD BSCCO compared to that of the optimally doped (OP) material. At $T = 295$ K, the spectra are similar, indicating that the power law region persists over a finite range of doping at high temperature. As the temperature is lowered, however, a gap-like feature appears in the OD spectra below 0.5 eV, indicating the emergence of an energy scale not present at optimal doping. This behavior is, at first glance, consistent with the emergence of a more Fermi liquid like phase at low temperature, and the presence of a fan-shaped quantum critical region centered on optimal doping (1,26,28,29).

The data do not, however, exhibit the generic properties of a quantum critical point (29). For one, the ~ 0.5 eV gap-like feature is more than an order of magnitude larger than the temperature scale on which it emerges, $k_B T$ ~ 20 meV. Furthermore, the M-EELS spectra are momentum-independent even in OD samples in which the gap-like feature is observed. Fig. 4 shows the momentum dependence of the OD data at $T = 115$ K (SI Appendix, Fig. S4 shows the data at 295 K). The spectra show very little $q$-dependence, just as in the OP case. We fit the data using Eqs. 1 and 2, though only for $\omega > 0.5$ eV (Fig. 4A), and obtain the fit parameters shown in Fig. 4B-D. Despite the clear appearance of an energy scale, the spectra still collapse (Fig. 4E), with $\Pi''(q,\omega)$ again exhibiting $q^2$ dependence below the crossover energy, $\omega_c(q)$ (Fig. 4B). Above the crossover energy (Fig. 4C), $\Pi''(q,\omega) \sim q^2 / \omega^2$ as in the OP case. We conclude that the momentum-independence of the susceptibility and the absence of an observable energy scale at optimal doping are unrelated effects with different physical origin.

We close by speculating about the underlying cause of the density fluctuations we observe. Our results bear a striking similarity to the so-called Marginal Fermi Liquid (MFL) hypothesis, which asserts that the strange metal is a consequence of a featureless continuum of fluctuations that pervades all time and length scales (30). This continuum is conjectured to arise from quantum fluctuations of some hidden order parameter that exhibits "local criticality," meaning that the spatial correlation length $\xi_x \sim \log \xi_t$, where $\xi_t$ is the temporal correlation length—a situation sometimes described as having a dynamical critical exponent $z = \infty$ (31). Our experiment affirms two aspects of the MFL hypothesis. The first is that a featureless continuum exists, and contains enough spectral weight to saturate the $f$ sum rule. The second is that, on energy scales less than $\omega_c$ ~ 1 eV or 10,000 K, the polarizability factors into independent functions of momentum and energy, $\Pi''(q,\omega) = f(q)g(\omega)$, which in MFL is the defining characteristic of local criticality, i.e., decoupling of space and time axes (31). We emphasize, however, that our experiment detects only charge fluctuations and provides no evidence either for or against the existence of loop currents. Furthermore, it would be important in future work to extend these measurements to lower energies, i.e. $\omega \leq k_B T$, to test the postulated $\omega/T$ prediction of the MFL polarizability.

Another possibility is that the response functions of the strange metal are dominated by disorder. BSCCO is known to be electronically inhomogeneous (32), and also exhibits an incommensurate supermodulation due to structural misfit between the $CuO_2$ and BiO layers (33). Disorder breaks translational symmetry and can explicitly broaden features in a momentum-resolved measurement such as M-EELS. Furthermore, random disorder has been shown, in simple spin models, to give rise to singular, frequency-independent correlation functions of the sort we observe here (34,35). Further studies of the response of strongly correlated systems to disorder are needed to clarify this issue.

Recently, theoretical approaches have been developed to address the strange metal problem from a completely new perspective. The anti-de Sitter/conformal field theory (AdS/CFT) correspondence, which relates a gravity theory in a curved spacetime to a strongly interacting quantum field theory on its boundary (36), is one such approach that has already been used to reproduce some properties of the strange metal holographically (37). Rapid developments in this area may shed new light on this problem.

In summary, we present the first *q*-resolved measurement of the dynamic charge susceptibility of a strange metal at the meV scale. We have uncovered a new type of charge dynamics in which the fluctuations are local to such a degree that space and time axes are effectively decoupled. Explaining this observation may require a new kind of theory of interacting matter.

## Acknowledgments

We gratefully acknowledge C. M. Varma for discussions and assistance analyzing the data, and J. Zaanen, N. D. Goldenfeld, and P. W. Phillips for helpful discussions. This work was supported by the Center for Emergent Superconductivity, an Energy Frontier Research Center funded by the U.S. Department of Energy, Office of Basic Energy Sciences under Award DE-AC02-98CH10886. Crystal growth was supported by DOE grant DE-SC0012704. P.A. acknowledges support from the EPiQS program of the Gordon and Betty Moore Foundation, grant GBMF4542. B.U. acknowledges NSF CAREER grant DMR-1352604. M.M. acknowledges support by the Alexander von Humboldt Foundation through the Feodor Lynen Fellowship program.

## Author Contributions

P.A. conceived of the project. S.V., A.K., and P.A. developed the M-EELS instrument. M.M., A.A.H., M.S.R. and S.I.R. performed the experiments. M.M. and A.A.H. analyzed the data with input from B.U. RPA calculations were performed by J. Schmalian. Samples were grown and characterized by R.Z., J. Schneeloch, and G.D.G. The manuscript was written by M.M. and P.A. with input from all authors.

## Author Information

The authors declare no competing financial interests.

## Methods

**Sample growth and characterization.** Optimally-doped single crystals of $Bi_{2.1}Sr_{1.9}Ca_{1.0}Cu_{2.0}O_{8+x}$ with $T_c$ = 91 K were grown by the floating-zone method (38). Overdoping was achieved by annealing in a hot, isostatic press with gas pressure 6.8 kbar at temperature 500 °C for 100h. The gas mixture was 20% O - 80% Ar with oxygen partial pressures up to 1.35 kbar. The $T_c$ values were determined using SQUID magnetometry.

**M-EELS measurements.**
M-EELS measurements were performed using an Ibach-type HR-EELS spectrometer (39) that was motorized and mated to a custom, multi-axis sample goniometer. Centering of the rotation axes was done using remote cameras and reference scatterers, as described previously (19). Experiments were done in a magnetically-shielded UHV chamber at $5\times10^{-11}$ torr vacuum and residual field of 3 mG using a 50 eV beam energy at 170 pA current and overall resolutions of 4 meV in energy and 0.02 Å$^{-1}$ in momentum. The BSCCO surfaces were prepared by cleaving along a (001) surface normal in a UHV prep chamber. Measurements were performed on three different optimally-doped and four overdoped crystals and tested on several cleaves of the same sample. Each figure of this manuscript reports data collected on a single, independent cleave. The crystals were oriented by locating elastic scattering from the (1,0) and (0,0) (specular) Bragg reflections and building an orientation matrix relating goniometer angles to momentum space (19). In this article, Miller indices $(H,K)$ indicate an in-plane momentum transfer $q = 2\pi(H,K)/a$, where $a = 3.81$ Å is the tetragonal lattice parameter. All measurements were carried out at a fixed out-of-plane momentum transfer of $L = 20$, defined in terms of a c-axis lattice parameter 30.8 Å$^{-1}$. Wide-energy scans were binned into 30 meV groups to improve statistics.

**Determining the susceptibility from M-EELS data.** The M-EELS cross section is given by (19,21)

$$\frac{\partial^2 \sigma}{\partial\Omega\partial\omega} = \sigma_0 M^2(q) \cdot S(q,\omega) \;, \tag{3}$$

where $\sigma_0 = I_0\sqrt{(E_i - \omega)/E_i}$ ($E_i$ being the incident electron energy) is a weakly energy-dependent prefactor and $S(q,\omega)$ is the dynamic structure factor of the surface. The Coulomb matrix element is given by

$$M(q) = \frac{4\pi e^2}{q^2 + (k_i^z + k_s^z)^2} \;, \tag{4}$$

where $q$ is the in-plane component of the momentum transfer, and $k_i^z$ and $k_s^z$ are the out-of-plane momenta of the incident and scattered electrons, respectively. As described previously (19), we obtained the density response, $\chi''(q,\omega)$, using the following procedure. First, $M^2(q)$ and $\sigma_0$ were divided from the experimental data, which yields $S(q,\omega)$ to within a multiplicative constant. We then removed the Bose factor by antisymmetrizing,

$$\chi''(q,\omega) = -\pi\left[S(q,\omega) - S(q,-\omega)\right] \;. \tag{5}$$

The overall scale was determined by applying the *f*-sum rule (19),

$$\int_0^{\omega^*} \omega\,\chi''(q,\omega)\,d\omega = -\pi N_{eff}(\omega^*)\frac{q^2}{2m} \;, \tag{6}$$

where *m* is the bare electron mass and $N_{eff} = 1.81 \times 10^{-4}$ Å$^{-3}$ is the effective carrier density derived from the optical loss function integral to ω* = 2.0 eV (23). In addition to providing an overall scale, the sum rule normalizes out systematic drifts in the beam intensity, changing beam footprint on the sample, etc.

**RPA Calculations.** The charge susceptibility, $\chi''(q,\omega)$, and polarizability, $\Pi''(q,\omega)$, reported in Fig. 1 are calculated within the random phase approximation (18) (RPA). The bare polarization bubble is determined by use of the Lindhard formula using a realistic tight-binding parametrization of the BSCCO band structure (40). The charge susceptibility is then determined through Eq. 1 using the Coulomb interaction for a layered electron gas, $V(q)$, for 2D layers separated by a constant *c* = 15.4 Å (41).

**Figures**

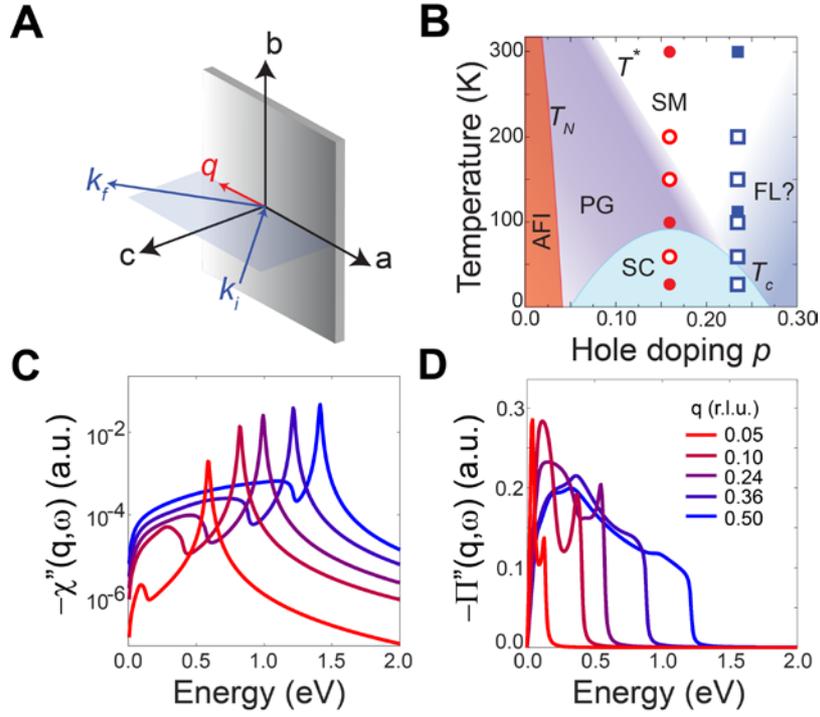

**Fig. 1 - Probing anomalous density fluctuations in the normal state of cuprates.** (A) Scattering geometry of the M-EELS experiment. $k_i$ and $k_f$ represent momenta of the incident and scattered electron, respectively, and $q$ is the in-plane momentum transfer. (B) Schematic temperature-doping phase diagram of BSCCO showing the points investigated in this work, with filled symbols indicating where a complete $q$-dependence was carried out. Here, AFI = Antiferromagnetic Insulator, PG = Pseudogap, SC = Superconductivity, FL = Fermi liquid, SM = Strange Metal, $T_N$ = Néel temperature, $T^*$ = Pseudogap temperature, $T_c$ = Superconducting critical temperature. (C) Charge susceptibility, $\chi''(q,\omega)$, of a layered electron gas calculated in the random phase approximation (RPA) using the Fermi surface parameterization of Ref. 40. (D) Associated charge polarizability $\Pi''(q,\omega)$.

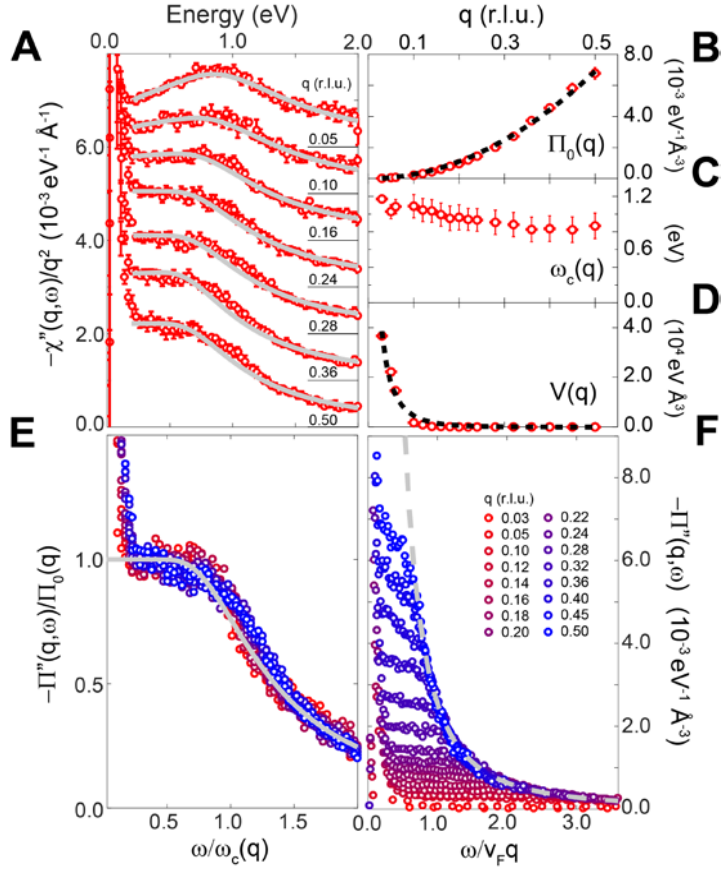

**Fig. 2 – Continuum collapse in optimally-doped BSCCO.** (A) Dynamic charge susceptibility, $\chi''(q,\omega)$, for a selection of momenta along the $(1,\bar{1})$ direction at 295 K (red symbols). The spectra were divided by $q^2$ and offset for clarity, the base line for each curve indicated by the solid line next to its momentum label. Error bars represent statistical, Poisson error. Grey lines are fits to the data using Eqs. 1 and 2 (see text). (B)-(D) Parameters used for the fits at every momentum measured (red symbols). $\Pi_0$ represents the overall magnitude of the continuum, $\omega_c$ is the crossover energy, and $V(q)$ is the Coulomb propagator near the surface. The dashed line in panel (B) represents a $q^2$ fit. The dashed line in panel (D) represents a fit using $V(q) \propto \exp[-qz]/q$ with $z = (8.1 \pm 1.5)$ Å. Errors in $q$ are given by the experimental momentum resolution. Parameter errors represent systematic uncertainty derived from a variation of $\pm 0.5$ in the exponent of the ratio $\omega_c(q)/\omega$ in Eq. 2. (E) Scaled collapse of the polarizability, $\Pi''(q,\omega)$, for all measured momenta (see text). The grey line is the fit function reported in Eq. 2. (F) Plot of the polarizability against the rescaled energy $\omega/v_F q$, showing $q^2/\omega^2$ behavior above the cutoff. The gray dashed line corresponds to $\Pi'' \propto q^2/\omega^2$.

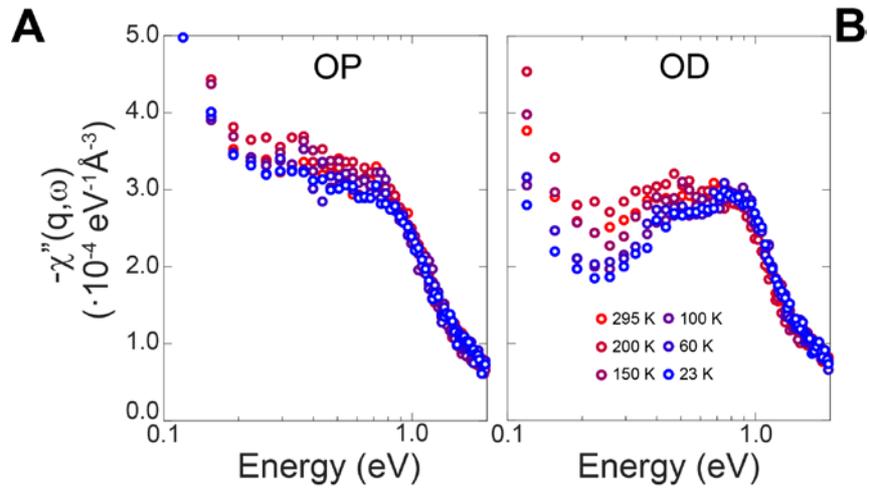

**Fig. 3 – Appearance of an emergent energy scale in overdoped BSCCO.** Temperature dependent $\chi''(q,\omega)$ at $q = 0.24$ r.l.u. along $(1,\bar{1})$ for (A) optimally-doped (OP) BSCCO and for (B) overdoped (OD) BSCCO. The OD data show the emergence of a ~ 0.5 eV energy scale as the temperature is lowered through the crossover region (Fig. 1A) (15).

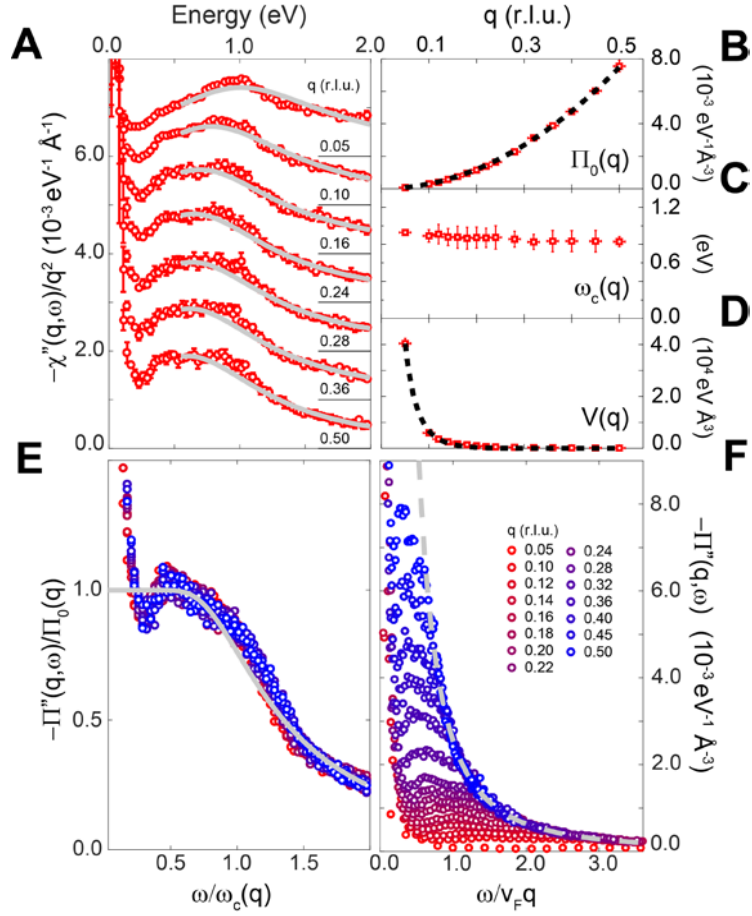

**Fig. 4 – Continuum collapse in overdoped BSCCO.** (A) Dynamic charge susceptibility, $\chi''(q,\omega)$, of overdoped BSCCO ($T_c$ = 50 K) for selected momenta along the $(1,\bar{1})$ direction at 115 K (red symbols). The spectra were divided by $q^2$ and offset for clarity. The spectra were divided by $q^2$ and offset for clarity, the base line for each curve indicated by the solid line next to its momentum label. Error bars represent statistical, Poisson error. Grey lines represent fits to the $\omega > 0.5$ eV region of the spectrum using Eqs. 1 and 2. (B)-(D) Parameters used for the fits at every momentum measured (red symbols). $\Pi_0$ represents the overall magnitude of the continuum, $\omega_c$ is the crossover energy, and $V(q)$ is the Coulomb propagator near the surface. The dashed line in panel (B) represent a $q^2$ fit. The dashed line in panel (D) represents a fit using $V(q) \propto \exp[-qz]/q$ with $z = (13.23 \pm 0.60)\,\text{Å}$. Errors in $q$ are given by the experimental momentum resolution. Parameter errors are the systematic uncertainty derived from a variation of $\pm 0.5$ in the exponent of the ratio $\omega_c(q)/\omega$ in Eq. 2. (E) Scaled collapse of the polarizability, $\Pi''(q,\omega)$, for all measured momenta (see text). The grey line from Fig. 2E is reproduced here for visual comparison to the gap-like

feature. (F) Plot of the polarizability against the rescaled energy $\omega/v_F q$, showing $q^2/\omega^2$ behavior above the cutoff. The gray dashed line corresponds to $\Pi'' \propto q^2/\omega^2$.

# Supporting Information for
# Anomalous density fluctuations in a strange metal

M. Mitrano[a,1], A. A. Husain[a], S. Vig[a], A. Kogar[a,b], M. S. Rak[a], S. I. Rubeck[a], J. Schmalian[c], B. Uchoa[d], J. Schneeloch[e], R. Zhong[e], G. D. Gu[e], P. Abbamonte[a,1]

[a] Department of Physics and Materials Research Laboratory, University of Illinois, Urbana, IL 61801, USA; [b] Department of Physics, Massachusetts Institute of Technology, Cambridge, MA 02139, USA; [c] Institute for the Theory of Condensed Matter, Karlsruhe Institute of Technology (KIT), 76131 Karlsruhe, Germany; [d] Department of Physics and Astronomy, University of Oklahoma, Norman, OK 73069, USA; and [e] Condensed Matter Physics and Materials Science Department, Brookhaven National Laboratory, Upton, NY 11973, USA.

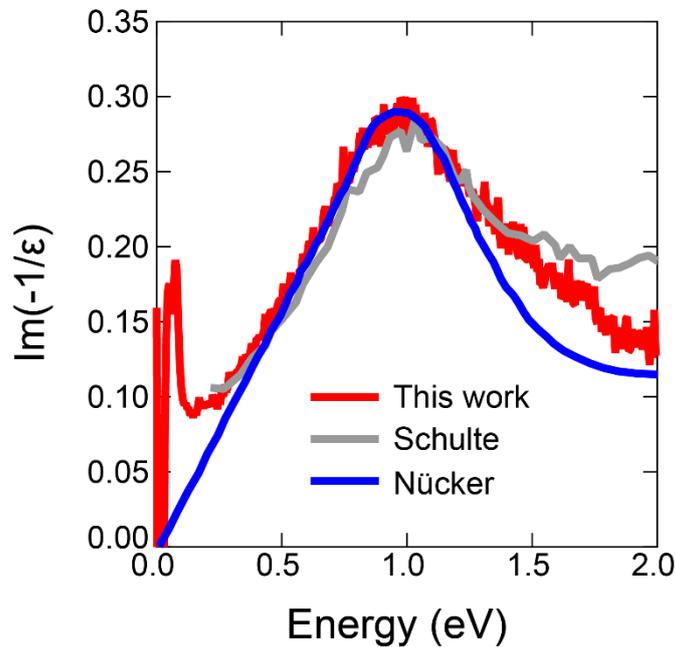

**Fig. S1 – Comparison with previous transmission and reflection EELS data.** Experimental loss function of BSCCO at 295 K along the (1,0) direction for q=0.05 r.l.u. (red) compared to a rescaled transmission EELS measurement for q=0.08 r.l.u. reproduced from Ref. 1 (blue) and to a rescaled reflection EELS spectrum at q=0.06 r.l.u. (grey) reproduced from Ref. 2.

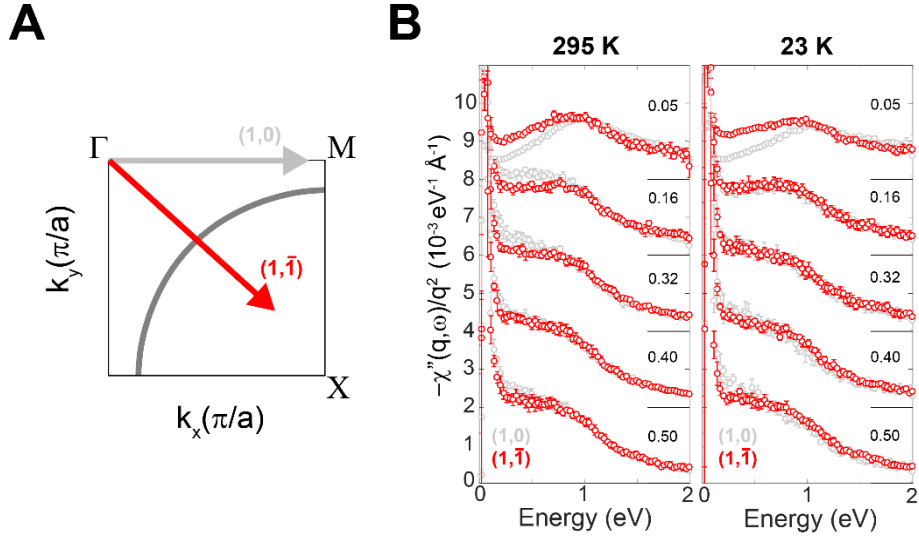

**Fig. S2 – Momentum space anisotropy of the χ"(q,ω) at optimal doping.** (A) Sketch of the 2D projected BSCCO Brillouin zone sector with the nodal direction orthogonal to the superlattice modulation. The red arrow indicates the $(1,\bar{1})$ direction, while the grey arrow is oriented along the antinodal $(1,0)$ direction. (B) Imaginary susceptibility χ"(q,ω) of optimally-doped BSCCO for selected momenta along the $(1,\bar{1})$ (red) and $(1,0)$ (grey) directions at 295 K (left) and 23 K (right). The spectra are normalized as described in the Methods, divided by $q^2$ and offset for clarity. Data are binned to 30 meV. Error bars represent statistical, Poisson error.

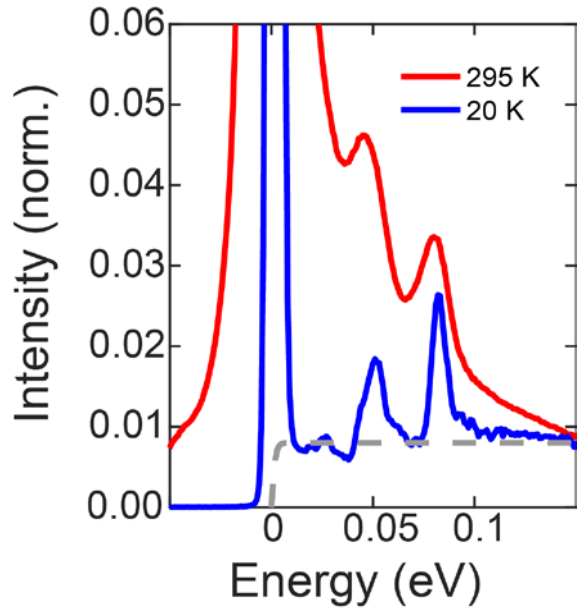

**Fig. S3 – High-resolution scans of optimally-doped BSCCO phonons.** Normalized EELS intensity at q=0 for $T$ = 295 K (red) and 20 K (blue). These data were acquired with 7 eV incident electron energy and 2 meV resolution. The dashed grey line is a low-energy extrapolation of the continuum discussed in the main text.

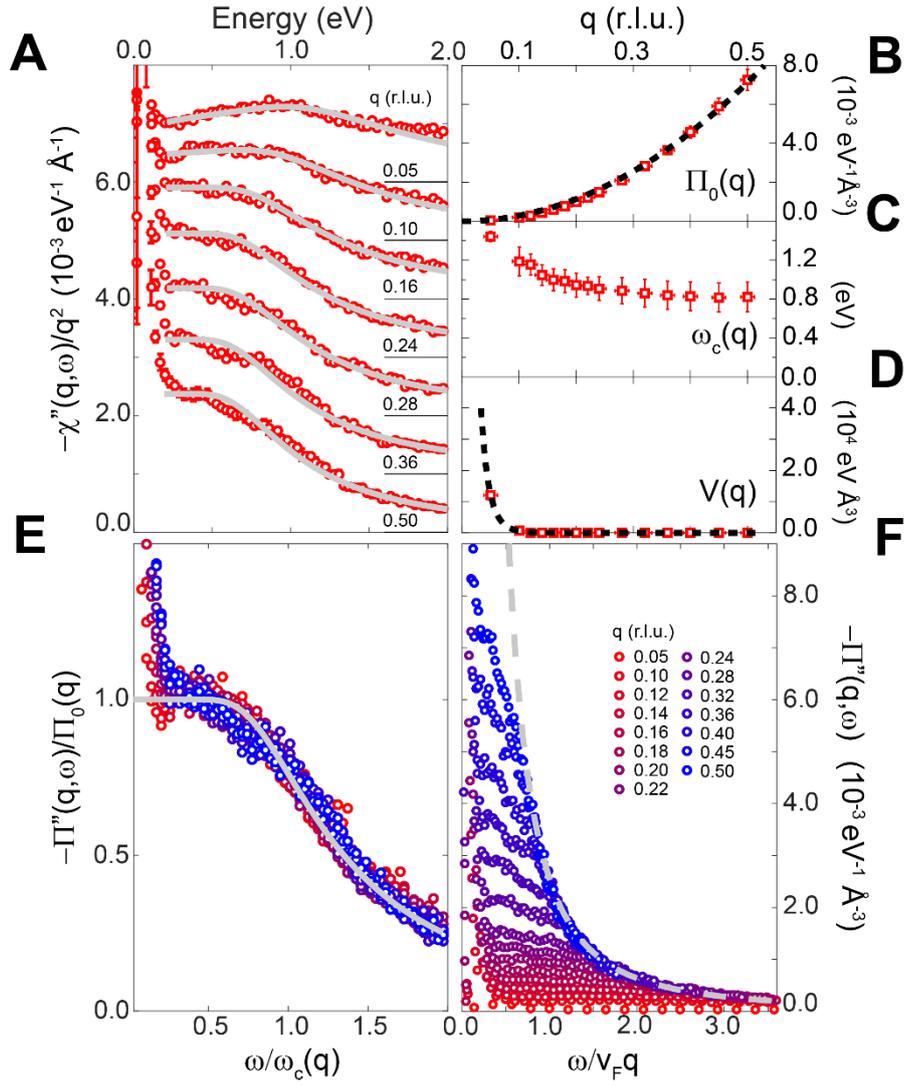

**Fig. S4 – Continuum collapse in overdoped BSCCO at 295 K.** (A) Dynamic charge susceptibility, $\chi''(q,\omega)$, of overdoped BSCCO ($T_c = 50$ K) for selected momenta along the $(1,\bar{1})$ direction at 295 K (red symbols). The spectra were divided by $q^2$ and offset for clarity. The base line for each curve is indicated by the solid line next to its momentum label. Error bars represent statistical, Poisson error. Grey lines represent fits to the data using Eqs. 1 and 2 of the main text. (B)-(D) Parameters used for the fits at every momentum measured (red symbols). $\Pi_0$ represents the overall magnitude of the continuum, $\omega_c$ is the crossover energy, and $V(q)$ is the Coulomb propagator near the surface. The dashed line in panel (B) represents a $q^2$ fit. The dashed line in panel (D) represents a fit using $V(q) \propto \exp[-qz]/q$ with $z = (28.2 \pm 1.2)\,\text{Å}$. Errors in $q$ are given by the experimental momentum resolution. Parameter errors represent systematic

uncertainty derived from a variation of $\pm 0.5$ in the exponent of the ratio $\omega_c(q)/\omega$ in Eq. 2. (E) Scaled collapse of the polarizability, $\Pi''(q,\omega)$, for all measured momenta (see text). The grey line is the fit function reported in Eq. 2. (F) Plot of the polarizability against the rescaled energy $\omega/v_F q$, showing $q^2/\omega^2$ behavior above the cutoff. The gray dashed line corresponds to $\Pi'' \propto q^2/\omega^2$.